\documentstyle [12pt,epsf]{article}

\begin{document}
%
%
\newcommand{\mod}[0]{\mbox{ mod }}
\newcommand{\Abs}[1]{|#1|}
\newcommand{\Tr}[0]{\mbox{Tr}}
\newcommand{\EqRef}[1]{(\ref{eqn:#1})}
\newcommand{\FigRef}[1]{fig.~\ref{fig:#1}}
\newcommand{\Abstract}[1]{\small
   \begin{quote}
      \noindent
      {\bf Abstract - }{#1}
   \end{quote}
    }
\newcommand{\FigCap}[2]{
\ \\
   \noindent
   Figure~#1:~#2
\\
   }
\newcommand{\beq}{\begin{equation}}
\newcommand{\eeq}{\end{equation}}
%
%
%
%
\title{Do zeta functions for 
intermittent maps have branch points?}
\author{
Per Dahlqvist \\
Mechanics Department \\
Royal Institute of Technology, S-100 44 Stockholm, Sweden\\[0.5cm]
}
\date{}
\maketitle

\ \\ \ \\

%
\Abstract{We present numerical evidence that the dynamical 
zeta function $\zeta(z)$ and the Fredholm
determinant $F(z)$ of intermittent maps with a 
neutral fix point  have branch point singularities at $z=1$.
We consider the power series expansion of $1/\zeta(z)$ and $F(z)$
around $z=0$
with the fix point pruned. This power series is computed up to order 20,
requiring $\sim 10^5$ periodic orbits.
We also discuss the relation between correlation decay and the nature of
the branch point.
We conclude by demonstrating how zeros of zeta functions
with thermodynamic weights that are
close to the branch point can be efficiently computed by a resummed cycle
expansion. The idea is quite similar to that of Pade\'{e} approximants,
but the ansatz is a generalized series expansion around the branch point
instead of a rational function.
}

\ \\

In this note we will present numerical evidence that dynamical zeta functions
and Fredholm determinants
exhibit branch point singularities for intermittent maps.
We will consider a family of one dimensional
maps $x \mapsto f(x)$ on the unit interval, with
\begin{equation}
f(x)=\left\{ 
\begin{array}{cc}
x+2^s x^{1+s} & 0\leq x <1/2\\
2x-1 & 1/2 \leq x \leq 1
\end{array}  \ \ ,
\right.
\end{equation}
where $s> 0$. For $s=0$ the map would just be the binary shift map,
which is uniformly hyperbolic, but for
$s>0$ it is intermittent; the fix point $x=0$ is neutrally
stable: $f'(0)=1$.
The map admits a binary coding, we associate the letter $0$ with the left leg,
and $1$ with the right leg. The neutral fix point now 
corresponds to the periodic orbit
$\overline{0}$.

The dynamical zeta function is defined by
\begin{equation}
1/\zeta(z)=\prod_p (1-\frac{z^{n_p}}{|\Lambda_p|})  \label{eqn:dyndef} \ \ .
\end{equation}
The product in \EqRef{dyndef}
runs over all primitive periodic orbits $p$, having period
$n_p$ and
stability $\Lambda_p=\frac{df^{n_p}}{dx}|_{x=x_p}$ with $x_p$ being any
point along $p$.

In our considerations we will prune the neutral fix point
\begin{equation}
1/\zeta(z)=(1-z)1/\tilde{\zeta}(z)  \  \  ,
\end{equation}
where
\begin{equation}
1/\tilde{\zeta}(z)=\prod_{p\neq \overline{0}} (1-\frac{z^{n_p}}{|\Lambda_p|})
\label{eqn:Zdef}  \ \ .
\end{equation}
We will also consider the Fredholm determinant
\begin{equation}
\tilde{F}(z)=\prod_{p\neq \overline{0}} 
\prod_{m=0}^{\infty}(1-\frac{z^{n_p}}{|\Lambda_p|\Lambda_p^m}) \ \ . \label{eqn:Fdef}
\end{equation}

In the previous case, it is not essential to prune the neutral
periodic orbit, but when considering the Fredholm determinant it is essential,
the extra factor $(1-z)^\infty$ would of course be devastating for
our investigations.

\vspace{0.3cm}


We will consider the series expansion of the functions $1/\tilde{\zeta}(z)$ and 
$\tilde{F}(z)$
\begin{equation}
1/\tilde{\zeta}(z)=1-\sum_{i=1}^\infty a_i z^i  \ \ , \label{eqn:a}
\end{equation}
\begin{equation}
\tilde{F}(z)=1-\sum_{i=1}^\infty b_i z^i  \ \ . \label{eqn:b}
\end{equation}

The nature of the 
leading singularity will be reflected in the asymptotic behaviour
of the coefficients of these power series. 
To get an idea what this asymptotic behaviour may be
we consider
the {\em fundamental part} of a {\em cycle expansion}
of $1/\tilde{\zeta}(z)$ \cite{AAC,ACL}
\begin{equation}
1/\tilde{\zeta}(z) \approx 1-\sum_{n=0}^{\infty} 
\frac{z^{n+1}}{\Lambda_{\overline{0^n1}}} \ \ .  \label{eqn:fund}
\end{equation}
The dependence of the stabilities $\Lambda_{\overline{0^n1}}$
on $n$ can be estimated by replacing the
difference equation $x_{n+1}=x_n+2^sx_n^{1+s}$ 
by the differential equation\cite{GT} 
\begin{equation}
\frac{dx_n}{dn}=2^sx_n^{1+s}  \ \ ,
\end{equation}
having solution
\begin{equation}
x_n = [x_0^{-s}-s2^s n]^{-1/s}  \  \  .
\end{equation}
First, because $x_n \sim 1$ we get a relation between $x_0$ and $n$
\begin{equation}
x_0 \sim n^{-1/s}  \  \  ,
\end{equation}
telling us how deep into the intermittent region 
the orbit $\overline{0^n1}$ penetrate.
We are only interested in the dependence on $n$ and have discarded other
information.
We can now obtain the stabilities by
differentiation
\begin{equation}
\Lambda_{\overline{0^n1}}=2\frac{dx_n}{dx_0}\approx
2x_n^{s+1}x_0^{-(s+1)} 
\sim n^{(s+1)/s}   \label{eqn:slope}  \  \ .
\end{equation}
This suggest that $\hat{Z}(z)$ contains a singularity of the type
\begin{equation}
\begin{array}{cc}
(1-z)^{1/s}  &  1/s \not\in N\\
(1-z)^{1/s}\log (1-z) & 1/s \in N^+  \ \ ,
\end{array}  \label{eqn:sing}
\end{equation}
as can be realized through the Tauberian theorems for power series.

A somewhat more refined, although closely related estimate is obtained by the
piecewise affine approximation of ref. \cite{Is}, leading to the same
predictions.
In \cite{Is} there are also much stronger arguments that $z=1$ 
is a branch point but
it is still  not proven that $1/\zeta(z)$ can be analytically continued outside
the unit disk. That this can actually be done is suggested by the numerical
data we are now going do discuss.

\vspace{0.3cm}

In order to 
numerically determine the coefficients $\{ a_i \}$ we simply expand the
product \EqRef{Zdef} using all periodic orbits up to period 20.
This set contains  
111011 periodic orbits which means that we cannot go very much further
within reasonable computer time.
When expanding the Fredholm determinant \EqRef{Fdef} we use Euler's formula
to expand the inner product:
\begin{equation}
\tilde{F}(z)=\prod_{p\neq \overline{0}} 
\prod_{m=0}^{\infty}(1-\frac{z^{n_p}}{|\Lambda_p|\Lambda_p^m})
\end{equation}
\[
=\prod_{p\neq \overline{0}}
\left(\sum_{j=0}^{\infty} (-1)^j\frac{\Lambda^{-\frac{j(j-1)}{2}}}
{|\Lambda_p|^j\prod_{k=1}^{j}(1-\Lambda^{-k})}z^{j\cdot n_p}\right) \  \  .
\]
Periodic orbits are determined by a Newton-Raphson procedure.
To this end we look for fix points of some iterate of the inverted map,
choosing branch according to the symbol code.

The coefficients for three parameter values are plotted in figs. 1-3
together with the expected slope according to eqs. \EqRef{fund} and
\EqRef{slope}. 
Our set of data is consistent with the expected 
leading singularity \EqRef{sing}
and indicate
no other singularity close to the unit circle; this would have induced
oscillations superposed on the sequence of coefficients.

For the binary shift map ($s=0$) we would have $a_i=1/2^i$.
For a slightly higher value of the parameter $s=0.1$ (fig. 1)
the coefficients for small $i$ conform with this behaviour but eventually
they bend off, approaching the expected slope.

One may also observe that whereas eq. \EqRef{fund} seems to provide the correct
order of the singularity it does not predict the correct size (prefactor) of it.
The curvature corrections of ref. \cite{AAC} exhibit the same type of
singularity as does the fundamental part.

\vspace{0.3cm}

Why is this interesting?
It is known that 
the spectra of zeta functions and Fredholm determinants
are, at least in some cases, related to the
(typical) decay of correlations. 
For exponentially mixing system
the mixing rate is given by
the size of the {\em gap} between the
leading and next-to-leading zeros of $F(z)$ \cite{Rue1}. 
We believe that a similar
coupling can be made system without a gap.
Let us take the (formal) trace of the transfer operator
\begin{equation}
tr {\cal L}^n=\int_{\epsilon}^{1} \delta (x-f^n(x))dx 
=\sum_{p\neq\overline{0}} n_p \sum_{r=1}^{\infty}  \frac{\delta_{n,rn_p}}
{\Abs{det(1-\Lambda_p^r)}}
=\frac{1}{2\pi i} \int_{|z|<1}
z^{-n}\frac{\tilde{F}'(z)}{\tilde{F}(z)}dz \label{eqn:integral} 
\end{equation}
where $\epsilon$ is a small number.
The trace is formal because it makes no explicit reference
to eigenvalues of an operator, it is just the trace over the integral
kernel of the operator.
We claim that this trace  serves as an archetype correlation function. 
If there is a gap, residue calculus tells us that 
the trace
will approach unity exponentially fast and the rate is provided by the size of
the gap.
But could this really work for the intermittent case 
where the leading zero is connected by a branch cut running along
the line $\mbox{Im} (z)=0$ $\mbox{Re} (z) >1$.
Let us assume that $\tilde{F}(z)$ is holomorphic and zero-free,
except along the cut, in the disk
$1<|z|<C$ where $C>1$. The value of the \EqRef{integral}
for large $n$
above will be governed by the vicinity of $z=1$  and can
be evaluated asymptotically.
Inserting $\hat{F}$ with the suggested branch point singularity
into \EqRef{integral} we get the following 
asymptotic behaviour for the trace
\begin{equation}
tr {\cal L}^n \sim \left\{ 
\begin{array}{ll}
1+C/n^{1/s-1} & 0<s<1\\
1+C/\log n & s=1\\
1/s & s<1 \end{array} \right.
\end{equation}
For $0<s<1$ this suggests that
typical correlations should decay as
$\sim 1/n^{1/s-1}$ which indeed agrees 
with the rigorous results \cite{Is}.
The failure of the trace to approach unity for $s>1$ reflect the fact that
the invariant density is not normalizable anymore.

There are indications that the identification between the behaviour of the
formal trace and the typical correlation functions is possible also for
the Sinai billiard which seems to exhibit the decay law $C(t) \sim 1/t$
\cite{PDsin,DAcorr}.

\vspace{0.3cm}

Our findings are also interesting from a more practical point of view.
Consider the problem of computing the topological pressure which amounts
to compute the leading zero of
\begin{equation}
1/\tilde{\zeta}(z,\beta)=\prod_{p\neq \overline{0}}
(1-\frac{z^{n_p}}{|\Lambda_p|^\beta}) \label{eqn:Ztp}  \ \ .
\end{equation}
If $\beta$ is close to, but less than unity this leading zero will be close
to $z=1$ and it will be extremely inefficient to compute it
from the truncated power series in $z$.
It is natural to try to expand $1/\tilde{\zeta}(z,\beta)$ in a generalized
power series around $z=1$. If the leading singularity is of the form
$(1-z)^\alpha$ the simplest possible expansion would be
\begin{equation}
1/\tilde{\zeta}(z,\beta)=\sum_{i=0}^{\infty} c_i (1-z)^i+
(1-z)^\alpha \sum_{i=0}^{\infty} d_i (1-z)^i \label{eqn:genseri} \ \ .
\end{equation}
According to our previous findings we expect that
$\alpha=\beta (s+1)/s-1$.
Suppose now that we replace these infinite sums by finite sums
of increasing degrees, $n_c$ and $n_d$, and require that
\begin{equation}
\sum_{i=0}^{n_c} c_i (1-z)^i+
(1-z)^\alpha \sum_{i=0}^{n_d} d_i
(1-z)^i=1/\tilde{\zeta}(z,\beta)+O(z^{n+1})  \label{eqn:genserf} \ \ .
\end{equation}
If $n_c +n_d=n+1$ we just get a linear system of equation to solve in
order to to determine the coefficients $c_i$ and $d_i$ from those of
the series
expansion around $z=0$. This strategy is quite similar to that of Pad\'{e}
approximants.
It also natural to require that $|n_d +\alpha -n_c|<1$.
So far we have assumed  that $\alpha$ is an non integer.
The case with integer $\alpha$ can be worked out in close analogy. 

To test the idea we choose (arbitrarily) the parameters $s=0.7$ and
$\beta=0.9$. In fig 4 we plot the leading zero versus truncation length
$n$ determined from expansion \EqRef{genserf} and the expansion around
$z=0$. The improvement is obvious. However, we do not claim
that the simple expansion \EqRef{genseri} is entirely able to capture the
complicate analytic structure around $z=1$. 

\vspace{0.3cm}

I would like to thank Viviane Baladi 
for gently persuading me to undertake this study.
This work was supported by the Swedish Natural Science
Research Council (NFR) under contract no. F-FU 06420-303.

\newpage

\newcommand{\PR}[1]{{Phys.\ Rep.}\/ {\bf #1}}
\newcommand{\PRL}[1]{{Phys.\ Rev.\ Lett.}\/ {\bf #1}}
\newcommand{\PRA}[1]{{Phys.\ Rev.\ A}\/ {\bf #1}}
\newcommand{\PRD}[1]{{Phys.\ Rev.\ D}\/ {\bf #1}}
\newcommand{\PRE}[1]{{Phys.\ Rev.\ E}\/ {\bf #1}}
\newcommand{\JPA}[1]{{J.\ Phys.\ A}\/ {\bf #1}}
\newcommand{\JPB}[1]{{J.\ Phys.\ B}\/ {\bf #1}}
\newcommand{\JCP}[1]{{J.\ Chem.\ Phys.}\/ {\bf #1}}
\newcommand{\JPC}[1]{{J.\ Phys.\ Chem.}\/ {\bf #1}}
\newcommand{\JMP}[1]{{J.\ Math.\ Phys.}\/ {\bf #1}}
\newcommand{\JSP}[1]{{J.\ Stat.\ Phys.}\/ {\bf #1}}
\newcommand{\AP}[1]{{Ann.\ Phys.}\/ {\bf #1}}
\newcommand{\PLB}[1]{{Phys.\ Lett.\ B}\/ {\bf #1}}
\newcommand{\PLA}[1]{{Phys.\ Lett.\ A}\/ {\bf #1}}
\newcommand{\PD}[1]{{Physica D}\/ {\bf #1}}
\newcommand{\NPB}[1]{{Nucl.\ Phys.\ B}\/ {\bf #1}}
\newcommand{\INCB}[1]{{Il Nuov.\ Cim.\ B}\/ {\bf #1}}
\newcommand{\JETP}[1]{{Sov.\ Phys.\ JETP}\/ {\bf #1}}
\newcommand{\JETPL}[1]{{JETP Lett.\ }\/ {\bf #1}}
\newcommand{\RMS}[1]{{Russ.\ Math.\ Surv.}\/ {\bf #1}}
\newcommand{\USSR}[1]{{Math.\ USSR.\ Sb.}\/ {\bf #1}}
\newcommand{\PST}[1]{{Phys.\ Scripta T}\/ {\bf #1}}
\newcommand{\CM}[1]{{Cont.\ Math.}\/ {\bf #1}}
\newcommand{\JMPA}[1]{{J.\ Math.\ Pure Appl.}\/ {\bf #1}}
\newcommand{\CMP}[1]{{Comm.\ Math.\ Phys.}\/ {\bf #1}}
\newcommand{\PRS}[1]{{Proc.\ R.\ Soc. Lond.\ A}\/ {\bf #1}}
%


\newpage

\begin{figure}
\epsffile{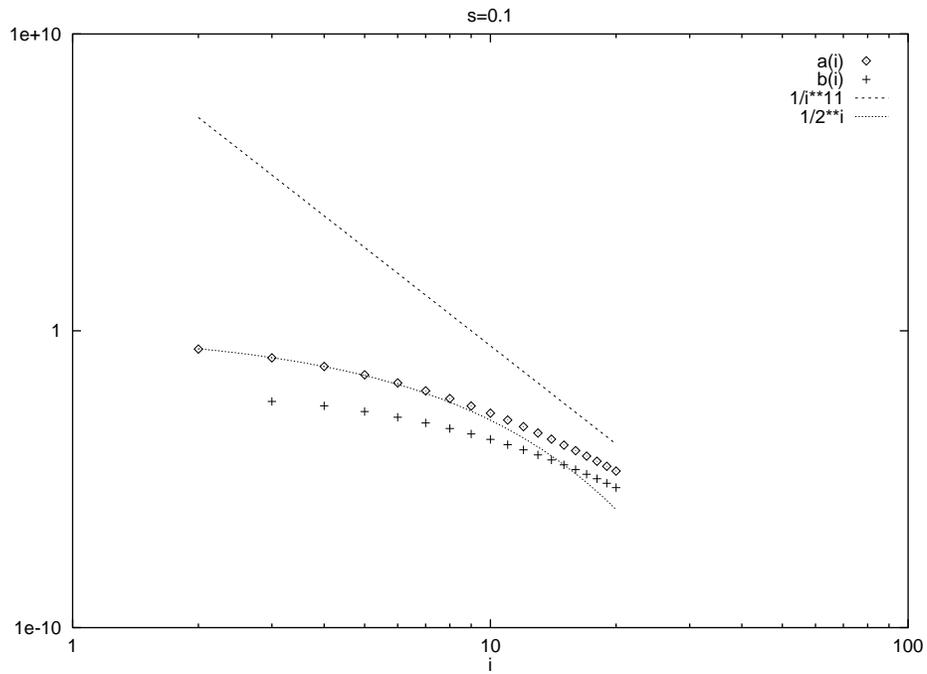}
\caption{Expansion coefficients for the functions $\hat{Z}(z)$ and
$\hat{F}(z)$ for the parameter values $s=0.1$} 
\end{figure}

\begin{figure}
\epsffile{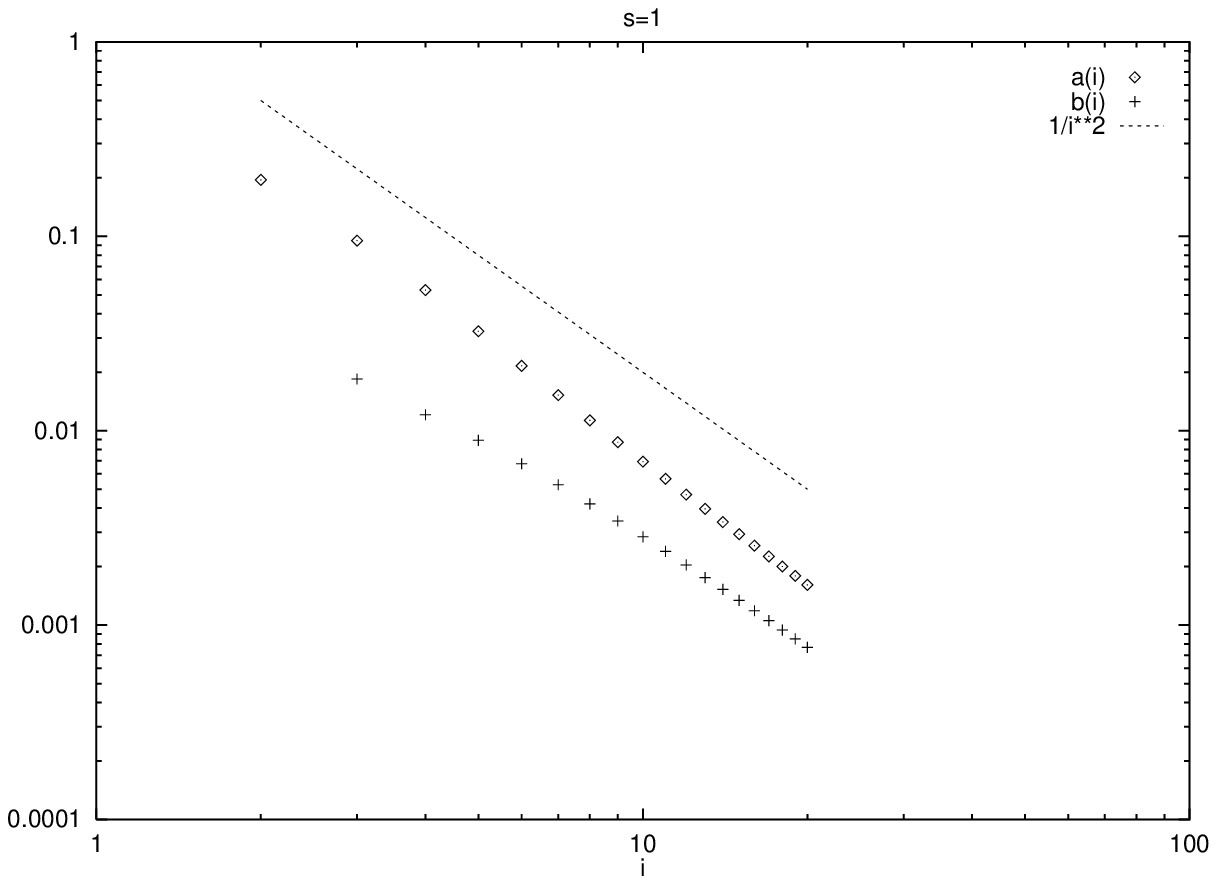}
\caption{Same as fig 1 but with $s=1$.} 
\end{figure}
 
\begin{figure}
\epsffile{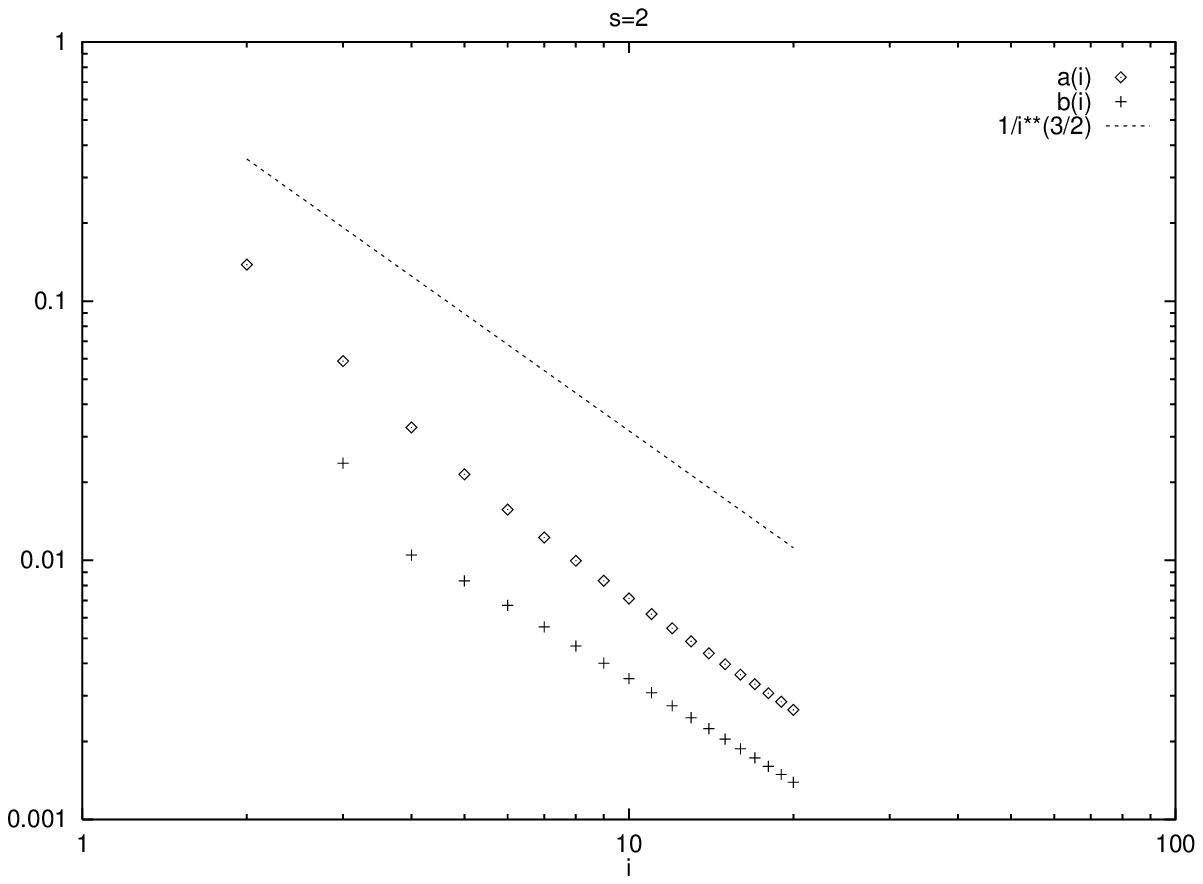}
\caption{Same as fig 1 but with $s=2$.} 
\end{figure}

\begin{figure}
\epsffile{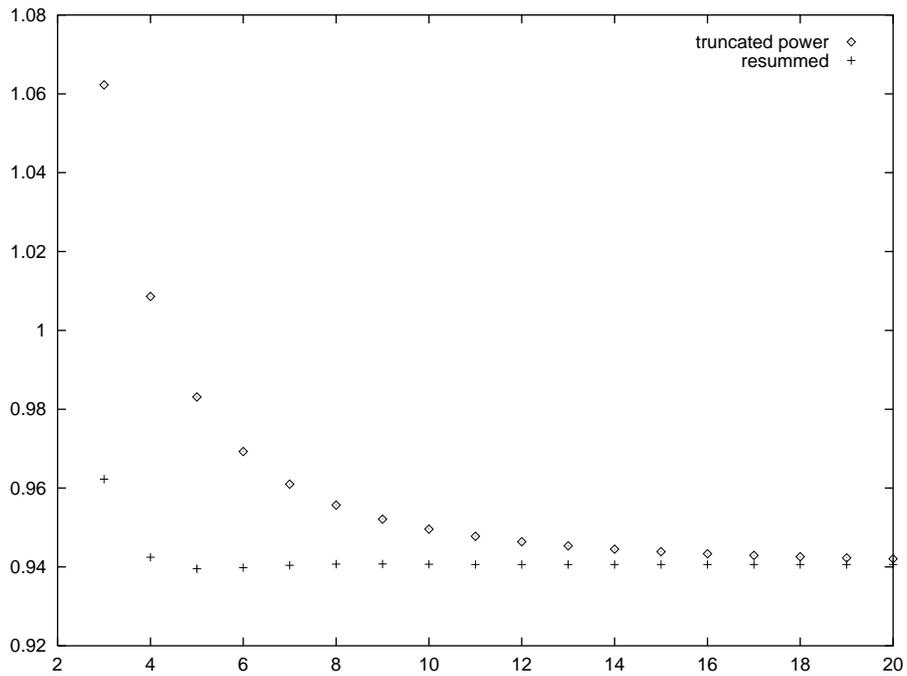}
\caption{Leading zero for $s=0.7$ and $\beta=0.9$
versus truncation length determined from the generalized series expansion
around $z=1$ and the power series around $z=0$} 
\end{figure}

\end{document}